%% file: conference_101719.tex
\documentclass{article}
\usepackage{spconf}

\usepackage{cite}
\usepackage{amsmath,amssymb,amsfonts}
\usepackage{algorithmic}
\usepackage[ruled,vlined, linesnumbered]{algorithm2e}
\usepackage{graphicx}
\usepackage{textcomp}
\usepackage{bm}
\usepackage{xcolor}
\def\BibTeX{{\rm B\kern-.05em{\sc i\kern-.025em b}\kern-.08em
    T\kern-.1667em\lower.7ex\hbox{E}\kern-.125emX}}

\include{macros_flinth}

\renewcommand{\vec}[1]{{\bf #1}}

\definecolor{bblue}{rgb}{0,0,0.8}

\newcommand{\sigmasi}{\sigma}

\title{Hierarchical sparse recovery from hierarchically structured measurements with application to massive random access}

\begin{document}

\name{Benedikt Gro\ss$^{1}$, Axel Flinth$^{2}$, Ingo Roth$^{3}$, Jens Eisert$^{3}$, Gerhard Wunder$^{1}$\thanks{This work was partially funded by DFG SPP 1798 - Compressed Sensing in Information Processing. F. acknowledges support from the Chalmers AI Research center (CHAIR).}}
\address{1. Department of Computer Science, 
Freie Universit\"at Berlin,
Berlin, Germany \\
2. Institution for Electrical Engineering, Chalmers University of Technology, 
Gothenburg, Sweden \\
3. Institute for Theoretical Physics,
Freie Universit\"at Berlin,
Berlin, Germany.}

\maketitle

\begin{abstract}
  A new family of operators, coined \emph{hierarchical measurement operators}, is introduced and discussed within the well-known hierarchical sparse recovery framework. Such operator is a composition of block and mixing operations and notably contains the Kronecker product as a special case. Results on their hierarchical restricted isometry property (HiRIP) are derived, generalizing prior work on recovery of hierarchically sparse signals from Kronecker-structured linear measurements. Specifically, these results show that, very surprisingly, sparsity properties of the block and mixing part can be traded against each other. The measurement structure is well-motivated by a massive random access channel design in communication engineering. Numerical evaluation of user detection rates demonstrate the huge benefit of the theoretical framework. 
\end{abstract}

\begin{keywords}
Structured compressed sensing, Hierarchically sparse signals, HiHTP, Block detection, Internet of Things (IoT), MIMO
\end{keywords}

\section{Introduction}
\subsection{The hierarchically sparse signal model}
The paradigm of compressed sensing \cite{candes:rip2008} can be boiled down to the notion that if a vector $\vec{x} \in \K^N$ ($\K$ refers to one of the two fields $\R$ or $\C$) is sparse, it can be recovered from underdetermined and noisy linear measurements $\vec{y}=\vec{A}\vec{x} + \eta \in \K^M$, $M \ll N$ using a plethora of algorithms.

The impact of compressed sensing on the signal processing cannot be understated. In many applications however, the signal $\vec{x}$ at hand is not only sparse, but enjoys a richer structure.
In this work, we consider \emph{hierarchically sparse} (hi-sparse) signals. Hi-sparse signals are structured into blocks that are sparse, or may even exhibit a block structure themselves. Sparsity is assumed on all levels of the block structure, i.e. in a hi-sparse signal $\vec{x} = (\vec{x}_1, \ldots, \vec{x}_N)$, only $s$ blocks $\vec{x}_i$ are non-zero, and each the non-zero blocks $\vec{x}_i$ contains at most $\sigma_i$ non-vanishing entries. 
Prior work on the recovery of \emph{block-sparse} signals includes \cite{EldarMishali:2009a,EldarMishali:2009b}, while a hi-sparse signal model has also been considered in
\cite{SprechmannEtAl:2010,FriedmanEtAl:2010, SprechmannEtAl:2011, SimonEtAl:2013}.
 Here, we will use the following definition. 

\begin{defi}
Let $\vec{x} = (\vec{x}_1, \dots, \vec{x}_N) \in \K^{n_1} \times \dots \times \K^{n_N}$, where $\K$ is either $\R$ or $\C$. For  $s$ and $\ {\bf\sigma}=(\sigma_1, \dots, \sigma_N)$, we say that $\vec{x}$ is $(s, \sigmasi)$-sparse, if
\begin{itemize}
    \item at most $s$ blocks $\vec{x}_i$ are non-zero and
    \item each non-zero block $\vec{x}_i$ is $\sigma_i$-sparse.
\end{itemize}
\end{defi}

Note that this definition can be readily generalized to more sparsity levels with a nested tree structure \cite{HiHTP} by allowing the $\sigma_i$ to be multi-leveled themselves, i.e. of the form $(\sigma_i, \bm{\varsigma}_i)$, with $\bm{\varsigma}_i=(\varsigma_{i,1}, \dots, \varsigma_{i,n_i})$ etc. 

Before formally introducing the new family of operators on hierarchically sparse signals, let us discuss a specific example in the context of massive random access in the IoT communication literature. For further application examples such as, e.g., massive MIMO please refer also to \cite{Wunder_ARX18},\cite{Wunder_TWC19}.

\subsection{Grouped Random Access Model}\label{sec:randomAccess}
In random access, each user that wishes to communicate with a base station chooses a resource at random (out of $\nu$ available) and sends a pilot signal associated with that resource. Assuming equal power transmitted from the users, and letting $\vec{b}_j(i) \in \C^m,m\leq\nu$, be the pilot signal that user $i$ chooses, the base station
%over $\nu$ time slots
then receives the vector
\begin{align*}
    \vec{y} = \vec{B}\vec{x} = \left(\sum_{i=1}^\nu x_i \vec{b}_{j(i)}\right),
\end{align*}
where $x_i =0$ if user $i$ is inactive, and $x_i=1$ else, and $\vec{B}=(\vec{b}_1,...,\vec{b}_{\nu})$. As before, it is reasonable to assume that the user activity is sparse meaning that the vector $\vec{x}$ is sparse.
%, say with at most $S$ non-zeros, and the base station must hence solve a sparse recovery problem to know which frequencies it must serve.

This protocol has a fundamental problem -- if several users choose the same resource (a collision occurs), subsequent communication is impossible, since each resource can only serve one user at a time. The probability of a collision grows very fast with a growing amount of users -- a phenomenon commonly referred to as the \emph{birthday paradox}.

\begin{figure}
    \centering
    \includegraphics[width=.45\textwidth]{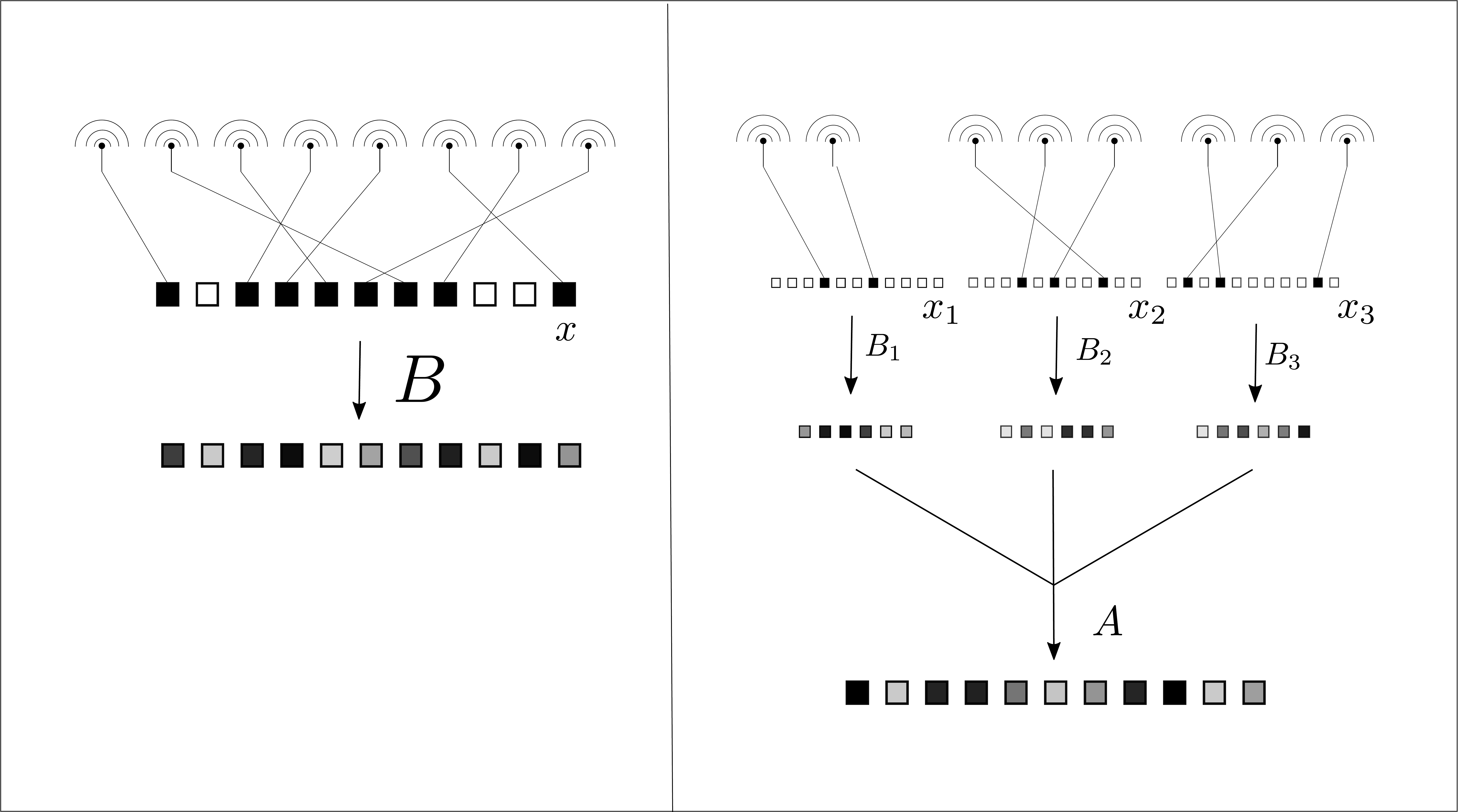}
    \caption{The standard random access (left) vs. the grouped approach proposed here (right).}
    \label{fig:user_detex}
\end{figure}

In order to reduce the probability of collisions, we propose to distribute the users in groups. That is, randomly subdivide the active users in $N$ groups, and accordingly replace the $x$-vector replaced with $N$ blocks $\vec{x}_i$. We then let the users in each group $i$ choose one of $n$ resources, with $n \leq \nu$. 
Obviously, there is then \emph{within each block} a much lower probability of collision than before. Furthermore, each block is with high probability sparse, say $\sigma_i$-sparse. Thus, the signal to be recovered, $(\vec{x}_1, \dots, \vec{x}_N) \in (\C^n)^N$ is no longer only sparse, but with high probability  even $(N,(\sigma_1, \dots, \sigma_N))$-hierarchically sparse. 

In order for the base station to be able to de-mix the individual block contributions, we may let them send their pilot signals during disjoint time intervals and recover each $\vec{x}_i$ from $\vec{B}_i\vec{x}_i$, for $i=1,...,N$ individually. This would however be very tedious compared to the original, one-shot scheme. Instead, we propose to mix the contributions $\vec{B}_i \vec{x}_i$ over (much smaller) $M$ incoherent slots (say in time, frequency or space), each time $j$ with a different random modulation $a_{j,i}$. The base station over the course of those time slots then receives
\begin{align}
    \vec{y}_j= \sum_{i=1}^N a_{j,i}\vec{B}_{i} \vec{x}_i, \quad j \in [M] \label{eq:groupedRandomAccess}
\end{align}
To further increase efficiency, we may let the $\vec{B}_i$ be subsampled, meaning that we have access to an $M\cdot m$-dimensional measurement, instead of an $N \cdot n$-dimensional. It is a priori not clear that recovery still is possible. However, the strict sparsity assumption of hierarchical sparsity still gives us some hope of recovery. We show that this intuition is indeed true and we bring a series of strong theoretical arguments why it is so.

\subsection{Hierarchical measurement operators}

The main aim of this paper is to analyze the properties of operators of the form  \eqref{eq:groupedRandomAccess} within the context of hierarchically sparse recovery. Let us give them a name.

\begin{defi}
	We call a measurement operator $\vec{H}$ from $\bigoplus_{i=1}^N\K^{n_i}$ to $\K^{M}\otimes \K^m$ a \emph{hierarchical measurement operator} if there exists matrices $\vec{B}_i \in \K^{m\times n_i}$ and a matrix $\vec{A} \in \K^{M\times N}$ with
	\begin{align}
		\vec{H}\left(\vec{x}_1, \dots, \vec{x}_N\right) = \sum_{i=1}^N \vec{a}_i \otimes (\vec{B}_i \vec{x}_i). \label{eq:hierarchical}
	\end{align}
\end{defi}
We will refer to $\vec{A}$ as the mixing matrix and the $\vec{B}_i$ as the block operators.

In \cite{HiHTP}, a subset of the authors proposed to use an adapted version of the celebrated Hard Threshold Pursuit (HTP) \cite{Foucart:2011} to recover hi-sparse signals -- the Hierarchical HTP (HiHTP), which we will present  in more detail in the experiment section.  The main finding of \cite{HiHTP} was that if a measurement operator $\vec{M}$ exhibits the so called \emph{hierarchically restricted isometry property (HiRIP)}, HiHTP recovers all hierarchically sparse signals in a stable and robust fashion from linear measurements $\vec{y} = \vec{M}\vec{x}$.

\begin{defi}
    The smallest  $\delta > 0$ for which
\begin{align*}
    (1-\delta) \norm{\vec{x}}^2 \leq \norm{\vec{M}\vec{x}}^2 \leq (1+\delta) \norm{ \vec{x}}^2 
\end{align*}
for all $(s,\sigma)$-sparse $\vec{x}$ is called the $(s,\sigma)$-HiRIP constant of $\vec{M}$, $\delta_{s,\sigma}(\vec{M})$.
\end{defi}

A matrix 'having the HiRIP' then refers to $\delta_{s,\sigma}(\vec{M})$ being small enough for the correct parameters $s$ and $\sigma$.

In this paper, we will analyse the HiRIP properties of hierarchical measurement operators.
We will give bounds on the HiRIP constants in terms of the RIP constants of the mixing and block matrices. These generalize prior work on compressed sensing of hierarchically sparse signals from Kronecker-structured measurements \cite{RothEtAl:2018}.

In the following, we will refrain from presenting any mathematical proofs. These, along with additional results and discussions, can instead found be found in the journal  version of this paper, of which a preprint \cite{Flinth2021hierarchical} will soon be available on arXiv.
\section{HiRIP-properties of Hierarchical Measurement Operators}

As briefly mentioned in the introduction, in \cite{RothEtAl:2018}, some of the authors of this article proved that a measurement matrix being a \emph{Kronecker product} $\mathbf A \otimes \mathbf B$
has the $(s,\sigma)$-RIP provided $\vec{A}$ has the $s$-RIP and $\vec{B}$ has the $\sigma$-RIP. 
The Kronecker product $\vec{A} \otimes \vec{B}$ is defined through its action on an element $(\vec{x}_1, \dots,  \vec{x}_N)\in (\K^n)^N$ as follows
\begin{align*}
    \vec{A}\otimes \vec{B}(\vec{x}_1, \dots,  \vec{x}_N) = \sum_{i=1}^N \vec{a}_i \otimes (\vec{B}\vec{x}_i),
\end{align*}
where $\vec a_i$ denotes the columns of $\mathbf A$, and $\otimes$ the tensor product. It is evident that these operators are special cases of the hierarchical operators considered here.

The main result of \cite{RothEtAl:2018} is that if $\vec{A}$ has the $s$-RIP and $\vec{B}$ has the $\sigma$-RIP, $\vec{A} \otimes \vec{B}$ has the $(s,(\sigma, \dots, \sigma))$-HiRIP. The first main result of this paper is a direct generalisation of this result.  

  \begin{theo} \label{th:HiRIP}
  	Let $\vec{H}$ be a hierarchical measurement operator, as in \eqref{eq:hierarchical}. Assume that the matrices $\vec{B}_i$ all obey the $\sigma_i$-RIP with constant $\delta_{\sigma_i}(\vec{B_i})$
  	 for all $i$. Assume further that $\vec{A}$ obeys the $s$-RIP with constant $\delta_s(\vec{A})$. Then $\vec{H}$ obeys the HiRIP, with
  	 \begin{align*}
  	 	\delta_{(s,\sigma)}(\vec{H}) \leq \delta_s(\vec{A}) + \sup_{i} \delta_{\sigma_i}(\vec{B}_i) + \delta_s(\vec{A}) \cdot \sup_{i} \delta_{\sigma_i}(\vec{B}_i),
  	 \end{align*}
  	 where again $\sigma=(\sigma_1,\dots, \sigma_N)$.
  \end{theo}
  
  \begin{rem}
    The result applies to more levels of sparsity. To be concrete, the $\sigma_i$ in the result may be multileveled themselves.
  \end{rem}
This result already enables us to construct an abstract hierarchical measurement operator which has the $(s,\sigma)$-HiRIP with high probability.  Remembering that a  Gaussian matrix $\vec{B}\in \K^{m,n_i}$ has $\sigma$-RIP with high probability already when  the number o
$ m \geqsim \sigma \log \left(\frac{n_i}{\sigma_i}\right)$\cite[Ch.9]{FouRau13}.
 In the same manner, we can choose the matrix $\vec{A} \in \K^{M \times N}$ as a Gaussian  matrix. It will have the $s$-RIP with high probability if $M \geqsim s\log \left(\tfrac{N}{s}\right)$. The above result then implies that the hierarchical measurement operator formed by $\vec{A}$ and the $(\vec{B}_i)_{i \in \N}$ has the $(s,\sigma)$-HiRIP. We conclude that recovery of $(s,\sigma)$-sparse vector is possible using asymptotically no more than
 \begin{align*}
    M \cdot m \geqsim s\max_i \sigma_i \cdot \mathrm{polylog}(N,n)
 \end{align*}
 total measurements, which is comparable to the results for unstructured measurement operators from \cite{HiHTP}. 
 
 \begin{rem}These considerations could to some extend already be derived using the Kronecker result from \cite{RothEtAl:iTwist:2016}. The main difference is that we may use varying sparsity levels, ambient dimensions and measurement operators.\end{rem}

\subsection{Refining the result through block operator incoherence}

As for the  Grouped Random Access model, Theorem \ref{th:HiRIP} does not bring a particularly satisfactory guarantee. Indeed, in that model, we have to recover $(N, \vec{\sigma})$-sparse signals. In order for the above result to be relevant, we thus need $\vec{A}$ to have the $N$-RIP, which necessitates $M \geq N$. This is discouraging.

Can the result however be strenghtened? In fact it can, if the block operators are incoherent in the following sense.

\begin{defi}
  We say that the collection of operators $\vec{B}_i \in \K^{m\times n_i}, i \in [n]$ are \emph{pairwise $(\delta,\sigmasi)$-incoherent} if for each $i\neq j$,
  \begin{align*}
      \sup_{\substack{|\vec{v}_i|_0\leq \sigma_i, |\vec{v}_j|_0\leq \sigma_j\\ 
      \norm{\vec{v}_i}= \norm{\vec{v}_j}=1}}
      \abs{\sprod{\vec{B}_i \vec{v}_i ,\vec{B}_j \vec{v}_j}}\leq \delta.
  \end{align*}
\end{defi}

In fact, the $\vec{B_i}$ in the Grouped Random Access is pairwise $(\delta,\sigmasi)$-incoherent if they are independently subsampled and modulated with random signs, as the following result shows.

\begin{prop} \label{prop:incoherentCollections}
         Let $\vec{F} \in \C^{n,n}$ be the unitary Fourier matrix.
        For each $i$, let the matrix $\vec{B}_i \in \C^{m,n}$ be formed by uniformly and independently sampling $m$ rows from $\vec{F}$, multiplying each of them with a uniform random sign, and subsequently rescaling the rows by $m^{-1/2}$. Assuming that $\vec{B}_i$ is independent of $\vec{B}_j$ for $i \neq j$ and
        \begin{align*}
            m \geqsim \sigma\delta^{-2} \log(n)^4\log(N),
        \end{align*}
         the collection is $(\delta,\sigma)$-incoherent with probability higher than $1-n^{-\log{n}^3}$.
\end{prop}

If  $(\vec{B}_i)_{i\in[N]}$ is a pairwise incoherent collection, it can to some extent separate the contributions of the individual blocks of a hierarchically sparse vector from the measurement $\sum_{i=1}^N \vec{B}_i \vec{g}_i$. Therefore, the mixing matrix $\vec{A}$ intuitively does not need to have a full $s$-RIP in order for the hierarchical measurement operator $\vec{H}$ formed by $\vec{A}$ and $(\vec{B}_i)_{i\in[N]}$ to have the $(s,\sigmasi)$-HiRIP. Put differently, if $\vec{A}$ is Gaussian, the needed number of $\vec{A}$ does not scale as $s$. A formal result is as follows.

\begin{theo}
\label{th:s_vs_delta}
    Let $\vec{B}_i \in \K^{m\times n_i}$ for $i\in[N]$, be a pairwise $(\delta_{2\sigma}^*,\sigma)$-incoherent family. Further assume that
    \begin{align*}
        \sup_{i} \delta_{\sigma}(\vec{B}_i) \leq \delta_\sigma^*
    \end{align*}
    Assume further that $(s\delta^*_{2\sigma})^2\leq N/\log(N)$ and that $\vec{A}\in \K^{M\times N}$ is a Gaussian matrix. Let furthermore $\delta,\epsilon>0$. Provided
    \begin{align*}
        M \geqsim \frac{(s\delta_{2\sigma}^*)^2}{\delta^2}\log\left(\frac{N(1+\delta_\sigma^*)^2}{(t\delta_{2\sigma}^*)^2}\right) + \log(\epsilon^{-1}),
    \end{align*}
    the hierarchical operator $\vec{H}$ defined by $\vec{A}$ and $(\vec{B}_i)_i$ obeys $
        \delta_{(s,\sigmasi)}(\vec{H}) \leq \delta + \delta_{\sigma}^*$
    with a probability at least $1-\epsilon$.
\end{theo}

The above result says that provided $\delta_{2\sigma}^*$ is small, we need $M$ to be of the order $(s\delta_{2\sigma}^*)^2$, rather than $s$, to allow for $s$-sparse signals on the 'block level'. If $\delta_{2\sigma}^*$ is very small, we may hence obtain the RIP already when $M$ is less than $s$ -- a behaviour which cannot be explained by Theorem \ref{th:HiRIP}. Note that we can even reach the realm of the  Grouped Random Access model, i.e. $s=N$.

\begin{rem}
    The quadratic dependence on of $s\delta_{2\sigma}^*$ is not sample optimal. We leave it to future work to determine whether this only is a proof artefact, or a fundamental limit. 
\end{rem}

  \section{Numerical Simulations}

In this section we empirically verify that the hierarchical sparsity framework, and in particular Theorem \ref{th:s_vs_delta}, can be used for user detection in the Grouped Random Access scenario.

\subsection{HiHTP Algorithm}
As mentioned in the introduction HiHTP is a low-complexity algorithm for solving hierarchically sparse compressed sensing problems of the form
\begin{equation}
\label{eq:problem}
    \min\limits_x \frac12\|y-\vec{H}x\|^2 \quad \text{ subject to $x$ is $(s,\sigmasi)$-sparse,} 
\end{equation}
where $\vec{H}$ is a linear operator satisfying the HiRIP. Provided $\vec{H}$ has the HiRIP for an appropriate sparsity level, it will recover any $(s,\sigma)$-sparse vector.

The algorithm consists of iteratively performing a gradient descent step, thresholding the new iterate onto the set of $(s, \sigmasi)$-sparse signals and then adjusting the non-zero values on the obtained support by linear least squares. The algorithm terminates once the support of two successive iterates does not change or a suitable stopping criterion is reached.

Importantly, the projection onto the set of $(s,\sigmasi)$-sparse signal (line 3 of Algorithm \ref{alg:hihtp}) can be performed in an efficient manner. We refer to \cite{HiHTP} for details on this.

\begin{algorithm}

\SetAlgoLined
\DontPrintSemicolon
\SetKwInOut{Input}{input}\SetKwInOut{Output}{output}\SetKwInOut{Init}{initialization}\SetKwFunction{Break}{break}
\Input{data $y$, stepsizes $\tau^{[t]}$, tolerance $\epsilon$}
\Init{$x^{(0)}=0$, $\Omega^{(0)}=\{\}$}      
\BlankLine
\For{$t=1, 2, \ldots$}{
$\hat{x}^{(t)} = x^{(t-1)} +\tau^{(t)} \vec{H}^*(y-\vec{H}x^{(t-1)})$\;
Calculate the support $\Omega^{(t)}$ of the best $(s,\sigmasi)$-sparse approximation of $\hat{x}^{(t)}$\;
$x^{(t)} = \argmin \frac12\|y-\vec{H}x\|^2$ subject to $\supp(x)\subset\Omega^{(t)}$\;
\If{$\Omega^{(t)}=\Omega^{(t-1)}$ {\bf or}  $\frac{\|y-\vec{H}x^{(t)}\|}{\|y\|}\leq \epsilon$}{
\Break}
} 
\caption{HiHTP for hierarchical measurements}
\label{alg:hihtp}
\end{algorithm}

\subsection{Grouped Random Access}%We demonstrate the implications of
We assume  $\nu=n=512$ available resources and model the measurements at the base station as $\vec{y} = \vec{F}\vec{x}$, where $\vec{F}\in\C^{n\times n}$ is a $n\times n$-DFT matrix. Note that since we are aiming for user detection, we do not necessarily need to recover $\vec{x}$ exactly: we only need to determine which $x_i$ are non-zero. Our baseline method therefore consists of computing $\supp(\vec{F}^{-1}y)$ to obtain the selected resources. 

\begin{figure}
    \centering
    \includegraphics[width=.49\textwidth]{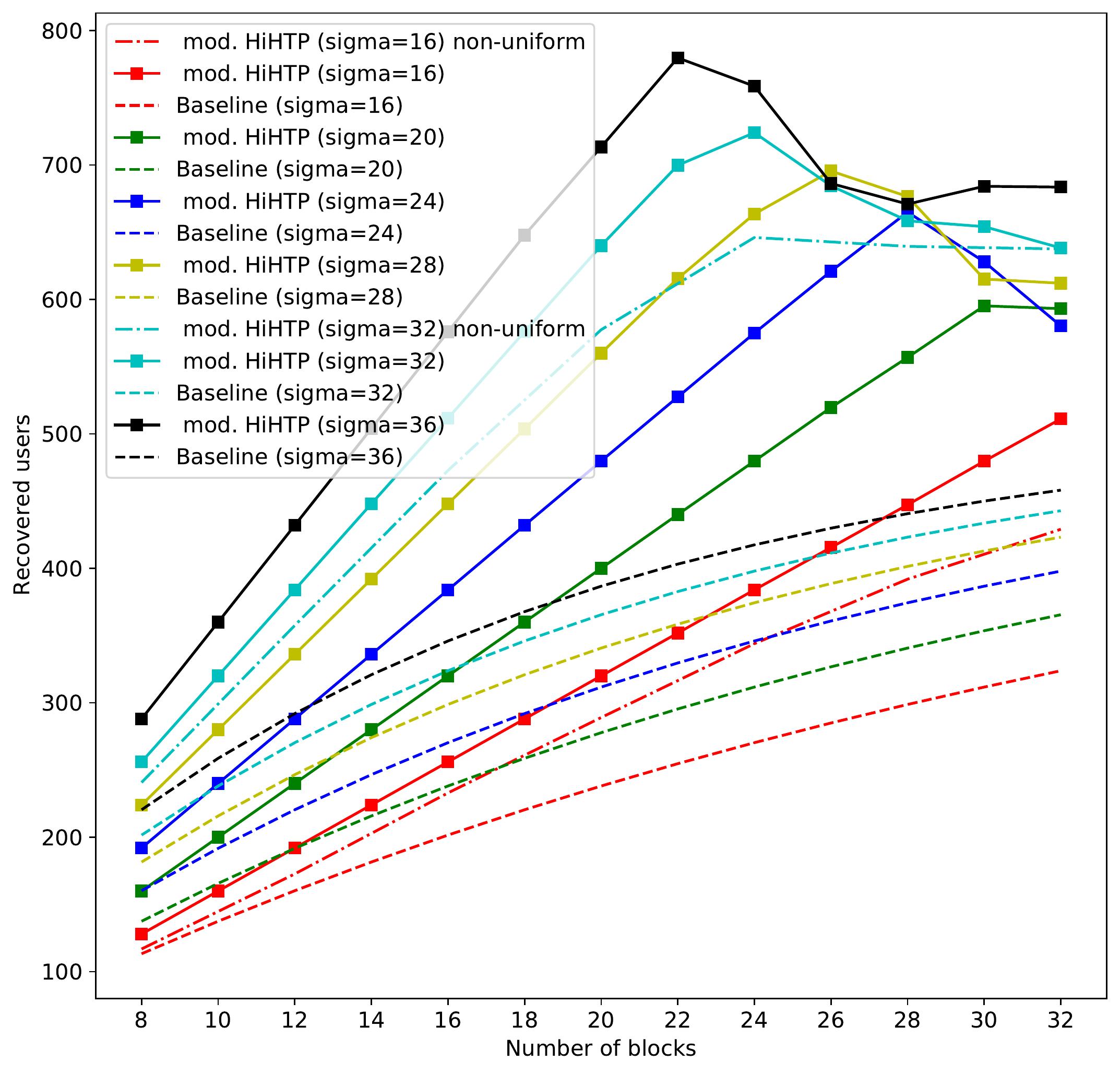}
    \caption{User detection with hierarchical measurements}
    \label{fig:user_detection}
\end{figure}

We compare this to the Grouped Random Access model with $N$ subsampled $\vec{B}_i$, which each consist of $m=256$ random rows from the $n\times n$ DFT matrix. The random modulation matrix $\vec{A}\in\C^{M\times N}$, where $M=16$ %is the number of antennas at the base station,
is chosen as complex Gaussian with variance $\tfrac{1}{\sqrt{N}}$. This results in the system $\C^{Mm}\ni\vec{y}= \vec{H}\vec{x}$, with measurement operator $\vec{H}$ formed by $\vec{A}$ and the $\vec{B}_i$ as in \eqref{eq:groupedRandomAccess}. 

The HiHTP Algorithm \ref{alg:hihtp} is used to recover $\vec{x}$ from the measurements $\vec{y}$. Figure \ref{fig:user_detection} shows the average number of recovered users for varying sparsities $\sigma=16,\dots,36$ and number of blocks $N=8,\dots,32$ over 25 Monte-Carlo trials for each configuration. Note that from $N>8$ on the system has more pre-image dimensions than measurements and the recovery results of our paper apply. Notably, since all the $N$ blocks are filled, classical hierarchical CS does not apply and particularly Theorem \ref{th:s_vs_delta} must be invoked. The baseline shown in Figure \ref{fig:user_detection} is computed as the total number of users, $N\cdot \sigma$, minus the expected number of collisions that occur, if these users choose randomly out of the $n=512$ available resources. As can be seen, HiHTP is able to recover much more users compared to the baseline. % Due to the distribution of the users into $N$ blocks, the average number of collisions in each block is $\coll(\sigma, n)<1$ and the system is able to serve much more users than the baseline method. 
For $\sigma=16$ and $\sigma=32$ the performance of the HiHTP algorithm for user detection is shown, when the $\sigma\cdot N$ users are distributed randomly over all available slots (i.e. the block sparsities are not uniformly fixed to $\sigma$). Even in this scenario, where the algorithm operates with the wrong sparsity parameters, reasonable performance is achieved.

  \section{Conclusion}
  We proved a number of results regarding the HiRIP for hierarchical measurement operators. The results generalize prior work on hierarchical sparse recovery with Kronecker-structured matrices. Furthermore, we demonstrated that the HiHTP algorithm is capable of computing hi-sparse solutions under this measurement structure in a practical Grouped Random Access design for IoT communication scenarios exhibiting huge capacity gains.

\bibliographystyle{unsrt}
\nocite{fantastic-5g-5g-ppp-project-on-5g-air-interface-below-6-ghz}
\bibliography{block_model,martin,ingo,gerhard}

\end{document}

%% file: macros_flinth.tex
\usepackage{hyperref}
\usepackage{verbatim} %to allow long "comment-out-'s"

\usepackage{graphicx}
\usepackage{caption}
\usepackage{subcaption}
\usepackage{amssymb}
\usepackage{ textcomp } %Promille sign
\usepackage{color}
\usepackage{enumerate}
\usepackage{mathrsfs}
\usepackage{multicol}
\usepackage{hyperref}
\usepackage{amsopn}

\usepackage{amsmath}
\usepackage{amsthm}
\usepackage{amssymb}

\usepackage{multicol}

\newcommand{\C}{\mathbb{C}}

\newcommand{\K}{\mathbb{K}}
\newcommand{\N}{\mathbb{N}}

\newcommand{\R}{\mathbb{R}}

\newcommand{\abs}[1]{\left\vert #1 \right\vert}
\newcommand{\norm}[1]{\Vert #1 \Vert}

\newcommand{\sprod}[1]{\left\langle #1 \right\rangle}

\usepackage{dsfont}

\newcommand{\geqsim}{\gtrsim}

\DeclareMathOperator{\supp}{supp}

\newcommand{\argmin}{\mathop{\mathrm{argmin}}}

\newtheorem{lem}{Lemma}
\newtheorem{prop}[lem]{Proposition}

\newtheorem{theo}[lem]{Theorem}

\newtheorem{defi}{Definition}
\newtheorem{rem}{Remark}

\theoremstyle{definition}

\numberwithin{lem}{section}

\usepackage{etoolbox}
\let\bbordermatrix\bordermatrix
\patchcmd{\bbordermatrix}{8.75}{4.75}{}{}
\patchcmd{\bbordermatrix}{\left(}{\left[}{}{}
\patchcmd{\bbordermatrix}{\right)}{\right]}{}{}